\documentstyle[12pt,epsfig]{article}
\newcommand{\ba}{\begin{eqnarray}} 
\newcommand{\ea}{\end{eqnarray}} 
\newcommand{\be}{\begin{equation}} 
\newcommand{\ee}{\end{equation}} 
\begin{document}
\thispagestyle{empty}
\title{Optimized perturbation method\\for the propagation
in the anharmonic oscillator potential}
\author{Anna Okopi\'nska\\ 
Institute of Physics, Bia\l ystok University,\\ 
Lipowa 41, 15-424 Bia\l ystok, Poland\\e-mail: okopin@fuw.edu.pl\\}
\maketitle
\begin{abstract}
\noindent The application of the optimized expansion for the
quantum-mechanical propagation in the anharmonic potential
$\lambda x^4$ is discussed for real and imaginary time. 
The first order results in the imaginary
time formalism provide approximations to the free energy and particle
density which agree well with the exact results in the whole
range of temperatures.
\end{abstract}
keywords: quantum mechanics, anharmonic oscillator, optimized
expansion, variational perturbation method, time evolution
amplitude, particle density
\clearpage
\noindent 1. The quantum mechanical Hamiltonian 
\begin{equation}
H=\frac{p^2}{2} +\frac{m^2 x^2}{2} +\lambda x^4
\end{equation}
provides the simplest representation of anharmonic effects and
is widely used in solid state physics, quantum chemistry and
even in paraxial optics (where the Helmholtz equation reduces to
the Schr\"odinger equation). Although the anharmonic oscillator
(AO) cannot be solved exactly, the energy spectrum, as well as
the time evolution amplitude can be calculated numerically to an
arbitrary accuracy. For systems with many degrees of freedom
such calculation become impractically time-consumming,
especially in the case of the time evolution, so that
approximation methods are desirable.

A very promissing and simple approximation method is generated
by the optimized expansion (OE). The method has been formulated
for the effective potential in quantum field theory in the
space-time of arbitrary dimension~\cite{AO}. Since the AO is
equivalent to the theory of a scalar field in the space-time of
one-dimension (time) with a classical action given by
\begin{equation}
{\cal A}[x]=\int_{0}^{T}\left[\frac{1}{2}(\dot x ^2(t)-m^2 x^2(t))-
\lambda x^{4}(t)\right]\,dt,
\label{Scl}
\end{equation}
the OE for the effective potential can be used to generate a
systematic approximation scheme for the free energy of the
system~\cite{AOT}, in the following called the optimized
expansion for the free energy (OEF). It has been shown that the
lowest orders of the OEF provide good approximations~\cite{AOT}
and the series converges to the exact free energy~\cite{conv}.\\

\noindent 2. Here we generalize the method to describe local
properties of the system, applying the OE to the evolution
amplitude. The real time propagator can be represented by
\be
(x_{b},T|x_{a},0)=\int_{x(0)=x_{a}}^{x(T)=x_{b}}
\!Dx\,e^{i {\cal A}[x]},
\label{K}
\ee
where the integral is taken over the functions which begin at
$x(0)=x_{a}$ and end at $x(T)=x_{b}$.

The most popular perturbation method is generated by expanding
the propagator~(\ref{K}) into a series in the coupling constant
$\lambda$ and performing the path integrals analytically, order
by order. Unfortunately the perturbation series for energy
eigenvalues of the AO is asymptotic~\cite{BW}, keeping thus a
few lowest terms in the expansion provides reasonable
approximation for very small values of $\lambda$ and short
propagation times $T$ only. Since the OE for the free energy
shows much better convergence properties than the perturbation
method, we expect also the approximations to the propagator
generated in the OE to have much broader region of
applicability.

The OE consists in modifying the classical action to the form
\begin{eqnarray}
{\cal A}^{mod}[x]=\int_{0}^{T}\left[\frac{1}{2}\left(\dot x^2(t)-
\omega^2 x^2(t)\right)+\epsilon\left(\frac{1}{2}(\omega^2-m^2)x^2(t)
-\lambda x^{4}(t)\right)\right]dt,
\label{Seps}
\end{eqnarray}
where the unperturbed part corresponds to the harmonic
oscillator with an arbitrary frequency, $\omega$. The formal
expansion parameter, $\epsilon$, has been introduced in such a
way that for $\epsilon=1$ the dependence on $\omega$ cancels and
the modified action becomes equal to the classical
one~(\ref{Scl}); therefore, after calculating the quantity of
interest to the given order in $\epsilon$ we set $\epsilon=1$.
The exact result, obtained as a sum of an infinite series, does
not depend on an arbitrary frequency, but a finite order
truncation does. The advantage of the freedom can be taken to
optimize the expansion: we make the $n$th-order approximant as
insensitive as possible to small variation of $\omega$, choosing
the value of $\omega$ which renders the approximant stationary.
The optimal frequency changes from order to order, which
improves convergence properties of the expansion scheme.

The OE for the time evolution can be generated by expanding
the amplitude~(\ref{K}) into a series
\ba
&&(x_{b},T|x_{a},0)=\int_{(x_{a},0)}^{(x_{b},T)}
\!Dx\,e^{i \int_{0}^{T}\left[\frac{1}{2}(\dot x^2(t)-\omega^2 x^2(t))
+\epsilon\left(\frac{1}{2}(\omega^2-m^2)x^2(t)-\lambda x^{4}(t)
\right)\right]dt}\nonumber\\&&=\int_{(x_{a},0)}^{(x_{b},T)}
\!Dx\,e^{i \int_{0}^{T} \frac{\dot x^2(t)-\omega^2 x^2(t)}{2}}
\left[1+i\epsilon\int_{0}^{T}\left((\omega^2-m^2)\frac{x^2(t)}{2}-\lambda x^{4}(t)
\right)dt+O(\epsilon^2)\right]\nonumber\\&&=\frac{\sqrt\omega}
{\sqrt(2\pi i\sin\omega T)} 
e^{\frac{i\omega\left[(x_{a}^2+x_{b}^2)
\cos\omega T-2 x_{a}x_{b}\right]}{2 \sin \omega T}}\left[1+
i\epsilon\frac{\omega^2-m^2}{2}\int_{0}^{T}[{\cal L}^2(t)+i{\cal K}(t)]dt\right.\nonumber\\&&\left.-
i\epsilon\lambda\int_{0}^{T} [{\cal L}^4(t)+6i {\cal L}^2(t){\cal K}(t)-3 {\cal K}^2(t)]dt+O(\epsilon^2)\right]~~~~~~~~~~~~~~~~~~~~~~~~~~~
\label{Kr}
\ea
where
\ba
{\cal L}(t)=\frac{x_{a}\sin\omega t+x_{b}\sin\omega (T-t)}{\sin\omega T}
{\mbox~~~~{\rm and}~~~~}
{\cal K}(t)=\frac{\sin \omega t\sin\omega (T-t)}{\omega\sin\omega T}.
\ea
Representing the evolution amplitude by
\be
(x_{b},T|x_{a},0)=e^{i {\cal W}(x_{b},x_{a},T)},
\label{Kw}
\ee
we calculate $W$ to $n$-th order in $\epsilon$ and upon setting
$\epsilon=1$ we optimize the approximant, ${\cal W}^{n}(x_{b},x_{a},T)$, 
choosing $\omega$ to fulfill
\ba
\frac{\delta {\cal W}^{n}(x_{a},x_{b},T)}{\delta\omega}=0.
\label{gap}
\ea
The optimal choice of $\omega$ in each order calculation,
ensures that the propagator about which we expand captures
essential features of the system under investigation.

The first order approximation is calculated to be given by
\ba
{\cal W}^{(1)}(x_{a},x_{b},T)&=&{\cal W}^{0}(x_{a},x_{b},T)
-i\frac{m^2-\omega^2}{2}\int_{0}^{T}[{\cal L}^2(t)+i{\cal K}(t)]dt\nonumber\\
&-&i\lambda\int_{0}^{T}[{\cal L}^4(t)+6i {\cal L}^2(t){\cal K}(t)-
3 {\cal K}^2(t)]dt,
\label{KGr}
\ea
where
\be
{\cal W}^{0}(x_{a},x_{b},T)=\frac{1}{2}\ln \left(\frac{\omega}
{2\pi i\sin\omega T}\right) + \frac{i\omega\left[(x_{a}^2+x_{b}^2)\cos\omega T-2 x_{a}x_{b}\right]}{2\sin \omega T}
~~~~~~
\label{K0}
\ee
corresponds to a harmonic oscillator with a frequency $\omega$.
The optimization condition~(\ref{gap}) reduces to
\be
\frac{m^2-\omega^2}{2}\int_{0}^{T}[{\cal L}^2(t)+i{\cal K}(t)]dt+
\lambda\int_{0}^{T} [{\cal L}^4(t)+6i {\cal L}^2(t){\cal K}(t)-3 {\cal K}^2(t)]dt=0
\label{gap0}
\ee
because of 
\ba
\frac{\delta {\cal W}^{0}}{\delta \omega^2}=
\frac{1}{2}\int_{0}^{T}[{\cal L}^2(t)+i{\cal K}(t)]dt.
\ea
For the harmonic oscillator with a frequency $m$ ($\lambda=0$)
the optimization condition~(\ref{gap0}) is solved by $\omega=m$,
so that the exact propagator is recovered in the discussed
approximation. Numerical results for the AO propagator
($\lambda\ne 0$) can be easily obtained, calculating however the
integral for wave function evolution
\be
\psi(x_{b},T)=\int (x_{b},T|x_{a},0)\psi(x_{a},0) dx_{a}
\ee
the problem of highly oscillatory behavior is encountered. We
decided therefore to discuss first the quality of the
approximation for the imaginary time propagator.\\

\noindent 3. The imaginary time evolution amplitude
\be
(x_{b},\beta|x_{a},0)=e^{W(x_{b},x_{a},\beta)}=
\int_{(x_{a},0)}^{(x_{b},\beta)}\!Dx\,e^{-A[x]},
\label{Ki}
\ee
where $\tau=it$ and the Euclidean action is given by 
\be
A[x]=\int_{0}^{\beta}\!\left[\frac{1}{2}(\dot x^2(\tau)+m^2 x^2(\tau))+
\lambda x^{4}(\tau)\right]\,d\tau,
\label{SE}
\ee
describes equilibrium properties of the system at temperature
$\beta^{-1}$. The trace of the imaginary time propagator defines
the partition function
\be
Z_{\beta}=\int\!dx_{a}(x_{a},\beta|x_{a},0)=\int\!dx_{a}\int_{(x_{a},0)}
^{(x_{a},\beta)}Dx\,e^{-A[x]}=\int\!dx_{a} e^{W(x_{a},\beta)}.
\label{Zt}
\ee
where $W(x_{a},\beta)=W(x_{a},x_{a},\beta)$ and the free energy
can be obtained as ${\mbox F_{\beta}=-\ln Z_{\beta}/\beta}$. The density
matrix can be expressed as
\be
\rho(x_{a},x_{b})=Z_{\beta}^{-1}\int_{(x_{b},0)}^{(x_{a},\beta)}
\!Dx\,e^{-A[x]},
\label{rom}
\ee
and the average particle density is given by its diagonal
element, $\rho(x_{a})=\rho(x_{a},x_{a})$.

The OE is generated by modifying the Euclidean action to the
form
\be
A^{mod}[x]=\int_{0}^{\beta}\!\left[\frac{1}{2}(\dot x^2(\tau)+
\omega ^2 x^2(\tau))+\epsilon\left[(m^2-\omega^2)x^2(\tau)+
\lambda x^{4}(\tau)\right]\right]\,d\tau.
\label{Amod}
\ee
The first order approximation for the imaginary time
amplitude~(\ref{Ki}), obtained by an analytic continuation of
the real time result~(\ref{KGr}), reads
\ba
W^{(1)}(x_{a},x_{b},\beta)=\frac{1}{2}\ln \left(\frac{\omega}
{2\pi \sinh\omega \beta}\right) +
\frac{\omega\left[(x_{a}^2+x_{b}^2)\cosh\omega \beta-2
x_{a}x_{b}\right]}{2\sinh \omega \beta}\nonumber\\
-\frac{m^2-\omega^2}{2}\int_{0}^{\beta}[L^2(t)+K(t)]dt
-\lambda\int_{0}^{\beta} [L^4(t)+6 L^2(t)K(t)+3K(t)^2]dt
\label{Wi}
\ea
where
\be
L(t)=\frac{x_{a}\sinh\omega t+x_{b}\sinh\omega (\beta-t)}{\sinh\omega \beta}
{\mbox{\rm~~~and~~~}}K(t)=\frac{\sinh \omega t\sinh\omega (\beta-t)}
{\omega\sinh \omega \beta}.
\ee

If we calculate the free energy by expanding the subintegral
expression in Eq.~\ref{Zt} to the given order in $\epsilon$ and
performing the Gaussian integrals over $x_{a}$ we would obtain
the series for the free energy 
\be
F_{\beta}=\frac{\omega}{2}+\frac{1}{\beta}\ln(1-e^{-\beta\omega})+
\epsilon\left[\frac{m^2-\omega^2}{2\omega}\left[
\frac{1}{2}+\frac{1}{e^{\beta\omega}-1}\right]+
\frac{3\lambda}{\omega^2}\left[
\frac{1}{2}+\frac{1}{e^{\beta\omega}-1}\right]^2\right] +O(\epsilon^2),
\label{EP}
\ee
which coincides with OEF~\cite{AOT}. The $n$-th order
approximation is obtained by requiring
\ba
\frac{\delta F^{n}_{\beta}}{\delta\omega}=0,
\label{gapF}
\ea
which determines $\omega$ as a function of $\beta$. The OEF provides
information on global properties of the system only. To discuss
local properties we proceed differently in this paper: we
perform the optimized expansion for the propagator (OEP),
imposing a local optimization condition
\ba
\frac{\delta W^{(n)}(x_{a},x_{b},\beta)}{\delta\omega}=0
\label{gapi}
\ea
which determines $\omega$ as a function of $\beta, x_{a}$ and
$x_{b}$. This approach yields an approximations to the density
matrix~(\ref{rom}) in a natural way. The given order
approximation to the partition function is obtained by performing 
integration over $x_{a}$ in Eq.~\ref{Zt} numerically and the
free energy is derived afterwards.

Here we discuss approximations to the free energy and to the
particle density for the one-dimensional AO, generated in the
first order of the OE. In Fig.~1 we show the results for the
quartic oscillator ($m=0,\lambda=1$): the free energy OEP
(obtained by optimization of the imaginary time amplitude) is
compared with OEF (obtained by optimization of the free energy)
and the exact result calculated numerically. In the limit of
high temperature both OEP and OEF approach the exact result, at
zero temperature the approximations coincide also, but the
accuracy is the worst. At finite temperature OEP appears
slightly better than OEF, differences between approximations are
not large oving to the fact that only a small region of $x_{a}$
contributes to the partition function and it does not make a big
difference whether the integration over $x_{a}$ is performed
before or after optimization. The quartic oscillator is on the
border between the single well $(m^2>0)$ and double well
$(m^2<0)$ AO. Since the accuracy improves for increasing
$\frac{m^2}{\lambda^{2/3}}$, the approximations to the free
energy for $m^2>0$ are better than in the case of quartic
oscillator. For the double well oscillator the accuracy is
worse, but even in this case both OEF and OEP are satisfactory
provided the wells are not very deep. The OEP has a great
advantage of yielding directly the approximations to the density
matrix~(\ref{rom}) and the particle density. The particle
densities, calculated in the first order of the OE for the
single well ($m^2=1$ and $\lambda=10$) and double well potential
($m^2=-1$ and $\lambda=.1$) are shown in Fig.2 and Fig.3,
respectively. They are in good agreement with the exact
densities calculated from the Schr\"odinger wave functions.

The OE for the imaginary time propagation amplitude bears some
similarities with the Feynman-Kleinert (FK) variational
method~\cite{GT,FK}, extended to a systematic variational
perturbation theory for the free energy~\cite{book}. In the FK
approach the exact partition function is expressed as
\begin{equation}
Z_{\beta}=\int\! \frac{dx_{0}}{\sqrt (2\pi\beta)}e^{-\beta V_{cl}(x_{0},\beta)},
\label{Vcl}
\end{equation}
with the classical effective potential, $V_{cl}(x_{0},\beta)$, defined by
\begin{equation}
e^{-\beta V_{cl}(x_{0},\beta)}= \int \! Dx\,\sqrt
(2\pi\beta )\delta(x_{0}-\overline{x}) e^{-A[x]},
\label{Zloc}
\end{equation} 
where $x_{0}=\overline{x}=\frac{\int\!d\tau x(\tau)}{\beta}$.
In the variational perturbation theory the same modified
action~(\ref{Amod}) is used, but the partition function is
obtained by performing numerical integration over $x_{0}$
in~(\ref{Vcl}) with $V_{cl}(x_{0},\beta)$ calculated to the
given order in $\epsilon$, i.e., the OE is applied to the
classical effective potential, while in our approach the
partition function is expressed by $W(x_{a},\beta)$ and
integration over $x_{a}$ in~(\ref{Zt}) is performed. The free
energy for the quartic oscillator obtained in the first order of
the FK approach is compared in Fig.1 with our results. In the
limits of high and low temperature the FK results coincide with
OEP, for intermediate temperatures the former agree better with
the exact free energy than the later. To any finite order in
$\epsilon$ the results of the methods are different, the
difference comes from the fact that $V_{cl}(x_{0},\beta)$ is a
function of the mean value of the coordinate, $x_{0}$, and
$W(x_{a},\beta)$ is a function of a starting point on periodic
trajectory, $x_{a}=x_{b}=x(t=0)$. The OEP can be thus directly
applied to calculate the particle density which is a function of
$x_{a}$. In the FK approach the relation between $x_{0}$ and
$x_{a}$ can be taken into account to calculate the particle
density~\cite{ro}, but this requires an additionary numerical
integration. The results of the FK method for particle densities which are
shown in Fig.2 and 3 are of similar accuracy as that obtained in
the OEP, the later are even better for the single well oscillator.
This is remarkable since the FK approach requires two
numerical integrations, while in the OEP only the integral for the
partition function has to be performed numerically, which is an
important advantage of our method in view of further
applications to the systems with many degrees of freedom.

It is worthwile to note that the FK method cannot be extended to
the case of non-periodic trajectories in order to obtain
non-diagonal terms of the density matrix and the propagation
amplitude in the real time formalizm. The OEP ofers such a
possibility in a natural way, we shall present a detailed
discussion of the real time propagation in a separate paper.
Both in the imaginary and in the real time formalism, the OEP
gives a possibility to calculate corrections to the propagation
amplitude, improving generated approximations in a systematic
way. In this way the convergence properties of the approximation
scheme can be estimated which is of great importance for the
systems where the exact result is difficult to obtain.\\

\noindent {\bf{\Large Acknowledgements}}\\

The author would like to thank K.~W\'odkiewicz to remind her
that local properties are important. The computational grant in
the Interdisciplinary Center of Mathematical Modelling of Warsaw
University is also acknowledged. This work has been supported
partially by the Comittee for Scientific Research under Grant
2-P03B-048-12.

\newpage
\noindent {\bf{\Large Figure captions}}\\

\noindent Figure 1. The free energy $F$ of the quartic
oscillator, $(m^2=0, \lambda=1)$ in the first order of the OE,
obtained by optimization of the imaginary time propagator (OEP,
{\it solid line}), and of the free energy (OEF, {\it
dashed-dotted line}) compared with the FK approximation ({\it
dashed line}) and the EXACT result ({\it dotted line}), 
plotted {\em vs.} the inverse temperature $\beta$.\\

\noindent Figure 2. The particle distribution of the single well
oscillator $(m^2=1, \lambda=10)$, obtained in the OE for the
imaginary time propagator (OEP, {\it solid line}), compared
with the FK approximation ({\it dashed line}) and the EXACT
result ({\it dotted line}) at $\beta=.1$ and $\beta=5$.\\

\noindent Figure 3. Same as in Fig.2, but for the double-well
oscillator ${\mbox (m^2=-1, \lambda=.1)}$ at $\beta=.25$ and
$\beta=5$.
\newpage
\begin{figure}
\epsfxsize=16cm\epsfbox{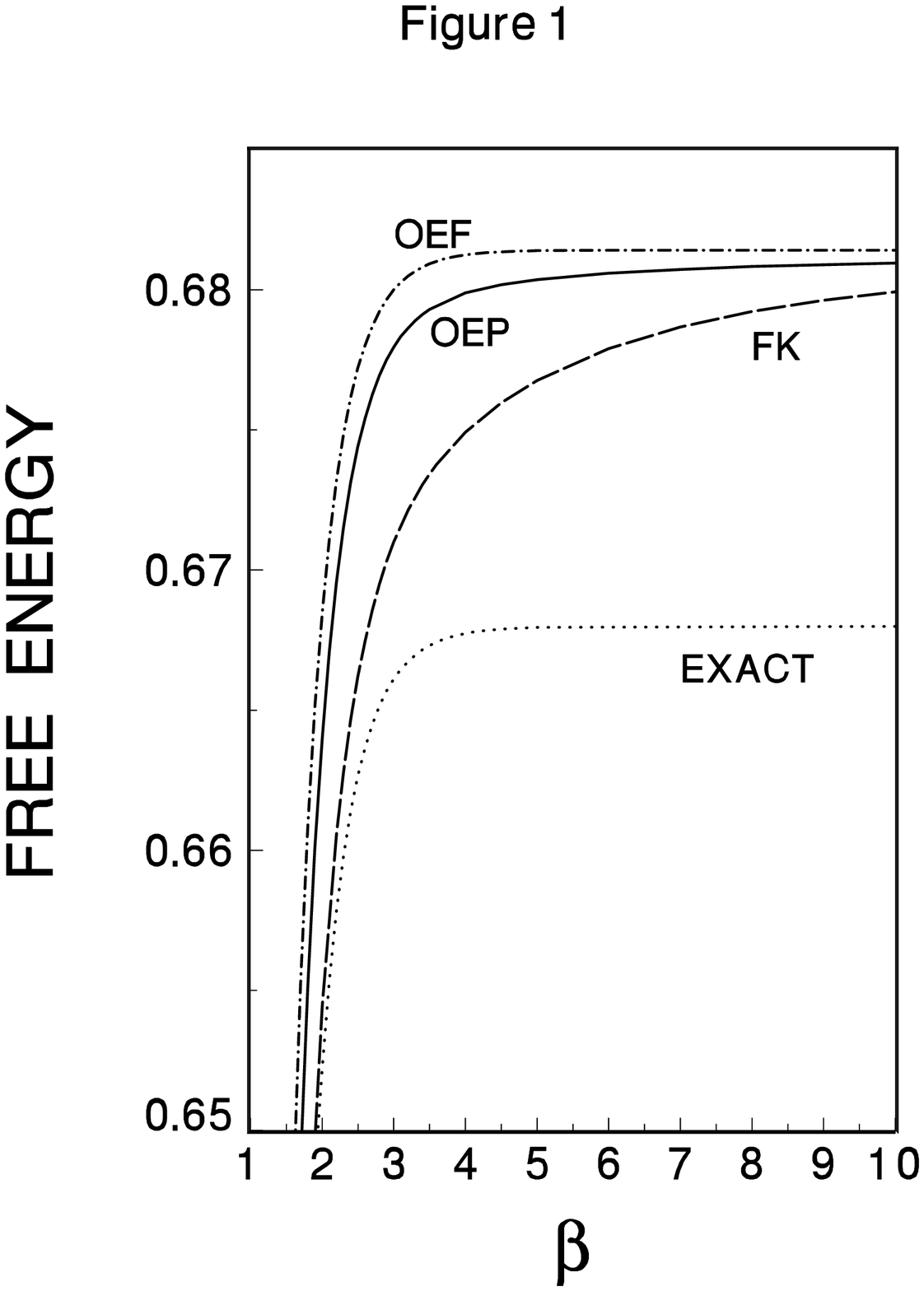}
\end{figure}
\newpage
\begin{figure}
\epsfxsize=16cm\epsfbox{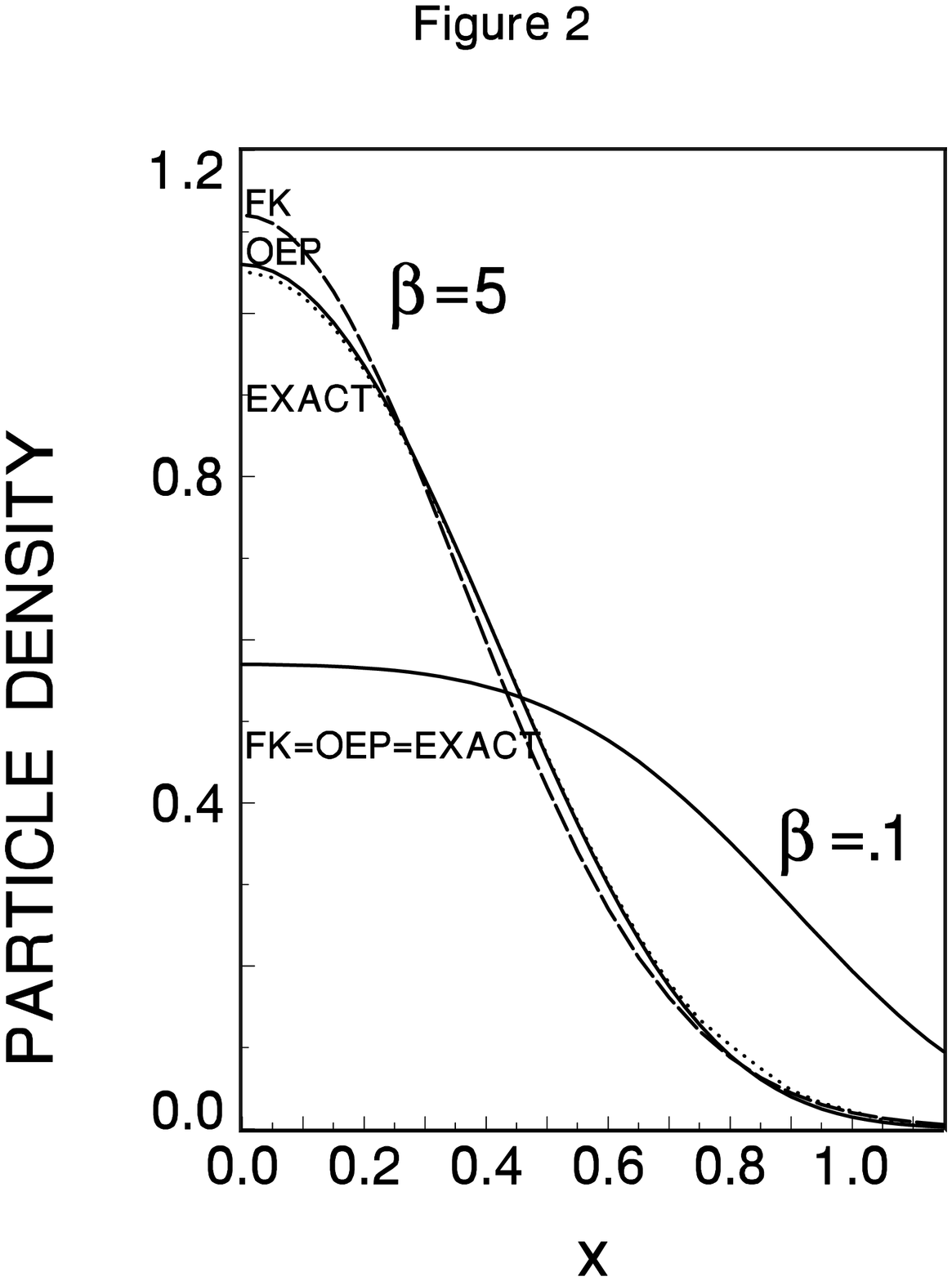}
\end{figure}
\newpage
\begin{figure}
\epsfxsize=16cm\epsfbox{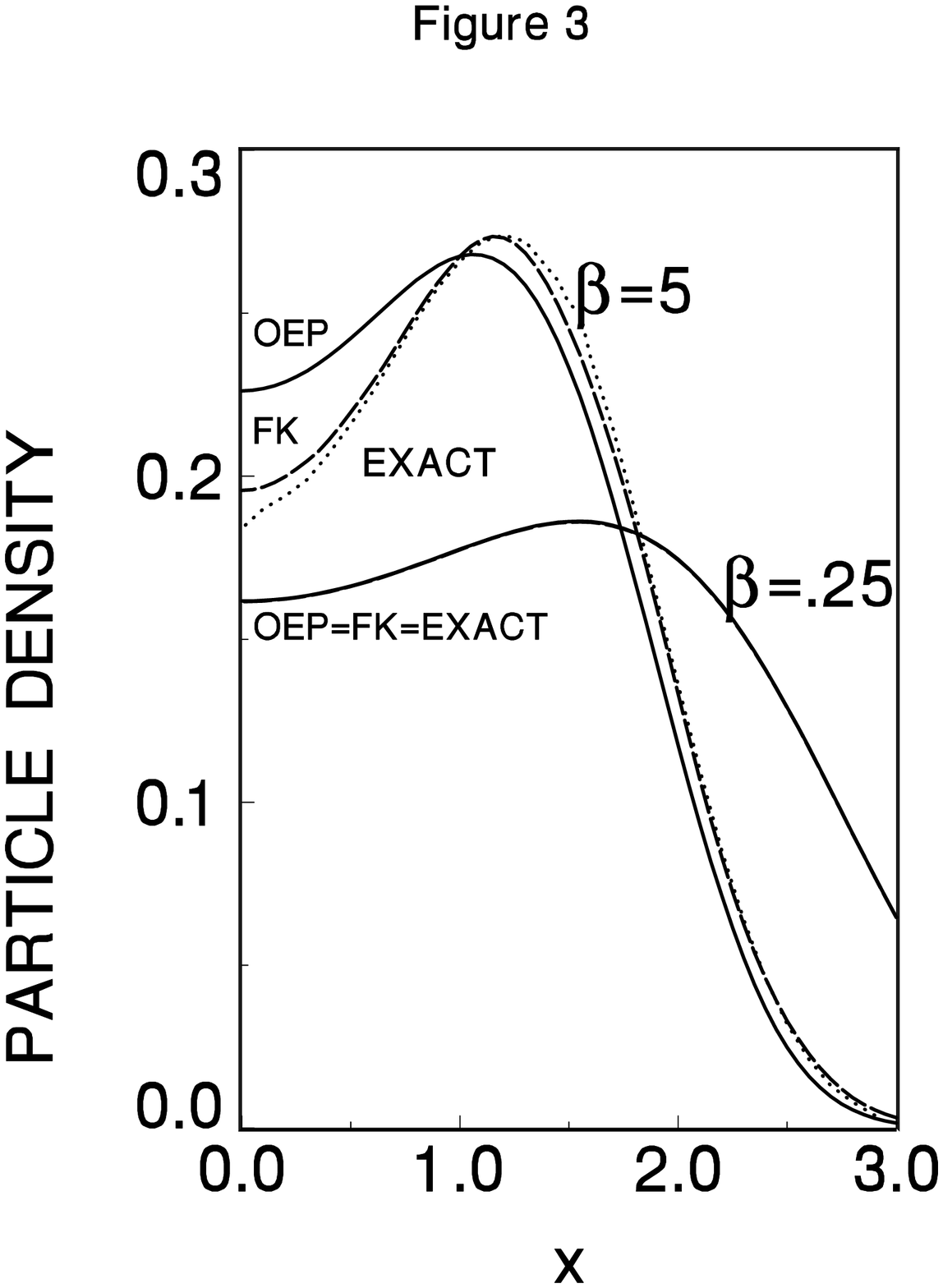}
\end{figure}
\end{document}